\documentclass[11pt]{article}
\usepackage{graphicx}

\newcommand{\BABARPubYear}    {05}

\newcommand{\BABARConfNumber} {008}
\newcommand{\SLACPubNumber} {11318}
\newcommand{\LANLNumber} {0000}

\input pubboard/babarsym

\newcommand{\xf}{\mbox{${\cal F}$}}
\newcommand{\costhr}{\ensuremath{\cos\theta_{\rm T}}}
\def\KS    {\ensuremath{K^0_{\scriptscriptstyle S}}}
\def\Bu      {\ensuremath{B^+}}
\def\Bub     {\ensuremath{B^-}}
\def\Bbar    {\overline{B}{}}

\def\Bzb     {\ensuremath{\Bbar^0}}
\def\Bz      {\ensuremath{B^0}}
\def\BpBm    {\ensuremath{\Bu  \Bub}}
\def\BzBzb   {\ensuremath{\Bz  \Bzb}}
\newcommand{\DE}{\ensuremath{\Delta E}}
\newcommand{\pvec}{{\bf p}}
\newcommand{\half}{\mbox{${1\over2}$}}
 \def\mes{\mbox{$m_{\rm ES}$}}
\newcommand{\calB}{\mbox{${\cal B}$}}
\newcommand{\auno}{\mbox{$a_1(1260)$}}
\newcommand{\aunob}{\mbox{$a_1$}}
\newcommand{\aunop}{\mbox{$a^+_1(1260)$}}

\newcommand{\appim}{\mbox{$a^+_1(1260) \, \pi^-$}}
\newcommand{\adueppim}{\mbox{$a^+_2(1320) \, \pi^-$}}
\newcommand{\pipppim}{\mbox{$\pi^+(1300) \, \pi^-$}}
\newcommand{\btoappim}{\mbox{$B^0 \rightarrow a^+_1(1260)\, \pi^-  $}}

\newcommand{\UfourS}{\mbox{$\Upsilon(4S)$}}
\newcommand{\Brapi}{\mbox{$\calB(\btoappim)$}}
\newcommand{\rapi}{\mbox{$40.2 \pm 3.9 \pm 3.9$}}
\newcommand{\Rapi}{\mbox{$(\rapi)\times 10^{-6}$}}
\def\BB{\mbox{$B\overline B\ $}}
\def\pep2{PEP-II}

\newcommand\etal{{\it et al.}}

\long\def\inst#1{\par\nobreak\kern 4pt\nobreak
    {\it #1}\par\vskip 10pt plus 3pt minus 3pt}

\begin{document}
{\pagestyle{empty}

\begin{flushright}
\babar-CONF-\BABARPubYear/\BABARConfNumber \\
SLAC-PUB-\SLACPubNumber \\
hep-ex/\LANLNumber \\
June 2005 \\
\end{flushright}

\par\vskip 5cm

\begin{center}
\Large \bf Measurement of the Branching Fraction  of $B^0$ Meson Decay to \appim
\end{center}
\bigskip

\bigskip \bigskip

\begin{center}
\large \bf Abstract
\end{center}

We present a preliminary  measurement of the branching fraction of the $B$ meson decay \\
\btoappim\   with \aunop\  $\rightarrow \pi^+ \pi^+ \pi^-$. The data sample corresponds to $218
\times 10^6$ \BB\ pairs  produced in \epem \ annihilation through the
\UfourS\ resonance.  We find the branching fraction $(40.2 \pm 3.9 \pm 3.9) \times 10^{-6}$,
where the first error quoted  is statistical and the second is systematic.
The fitted values of the \auno\ parameters are $m_{\aunob }=1.22 \pm 0.02$ \gevcc\ and
$\Gamma_{\aunob } = 0.423 \pm 0.050$ \gevcc.

\vfill
\begin{center}

Contributed to the
XXII$^{\rm st}$ International Symposium on Lepton and Photon Interactions at High~Energies, 6/30 --- 7/5/2005, Uppsala, Sweden

\end{center}

\vspace{1.0cm}
\begin{center}
{\em Stanford Linear Accelerator Center, Stanford University, 
Stanford, CA 94309} \\ \vspace{0.1cm}\hrule\vspace{0.1cm}
Work supported in part by Department of Energy contract DE-AC03-76SF00515.
\end{center}

\newpage
} 

\begin{center}
\small

The \babar\ Collaboration,
\bigskip

B.~Aubert,
R.~Barate,
D.~Boutigny,
F.~Couderc,
Y.~Karyotakis,
J.~P.~Lees,
V.~Poireau,
V.~Tisserand,
A.~Zghiche
\inst{Laboratoire de Physique des Particules, F-74941 Annecy-le-Vieux, France }
E.~Grauges
\inst{IFAE, Universitat Autonoma de Barcelona, E-08193 Bellaterra, Barcelona, Spain }
% ...

\end{center}\newpage

\section{INTRODUCTION}

We report on the preliminary  measurement of  the branching fraction
$B^0 \rightarrow  a_1^+(1260) \pi^-$ with
$a_1^+(1260) \rightarrow \pi^+ \pi^+ \pi^-$\cite{Coniugati} .
The  $a_1(1260) \rightarrow 3 \pi$
decay  proceeds mainly through the intermediate states
$(\pi \pi)_{\rho} \pi$ and $(\pi \pi)_{\sigma} \pi$~\cite{PDG2004} .

The study of this decay mode is complicated by open questions on the
parameters of the $a_1(1260)$ meson. There are large discrepancies between
 these parameters when comparing results from analyses involving hadronic 
interactions~\cite{palombo} and $\tau$ decays~\cite{tau}. Therefore, it is important
to verify the theoretical prediction of the branching fraction for
this decay mode and have  new measurements of the $a_1(1260)$ parameters. 
A theoretical calculation of the branching fraction of this decay mode
has been made by Bauer, Stech and Wirbel (BSW)~\cite{Bauer}
within the framework of the factorisation model. They predict a value of
$38 \times 10^{-6}$, assuming $\left|\frac{V_{ub}}{V_{cb}}\right|$ = 0.08.
It is also important to note that the $B^0 \rightarrow  a_1^+(1260) \pi^-$ channel 
can be used to measure the Cabibbo-Kobayashi-Maskawa
angle $\alpha$ of the Unitarity triangle~\cite{aleksan}. We presented a preliminary 
version of this analysis at ICHEP'04~\cite{ICHEP04}, using an integrated luminosity 
of $112 fb^{-1}$ and the measured branching fraction 
was $(42.6 \pm 4.2 \pm 4.1)$$\times 10^{-6}$.
For the branching fraction of $B^0 \rightarrow a_1^+(1260) \pi^-$ an upper limit 
of $49 \times 10^{-5}$ at the 90\% confidence level (C.L.) has been set by CLEO 
collaboration~\cite{CLEO} while the DELPHI collaboration~\cite{Delphi}
has set the 90\% C.L. upper limit of $28 \times 10^{-5}$ for the
branching fraction of $B^0 \rightarrow 4\pi$.

Below we present the details of the analysis for the measurement of the
branching fraction for $B^0 \rightarrow  a_1^+(1260) \pi^- \rightarrow 2\pi^+ 2\pi^-$.
Presently, we do not distinguish between the final
states $(\pi \pi)_{\rho} \pi$ and $(\pi \pi)_{\sigma} \pi$. Such an analysis
would require a study of the angular distributions of the decay products. 
Possible background contributions from $B^0$ decays to $a_2^+(1320) \pi^-$ and $\pi^+(1300) \pi^-$
are studied and taken into account while in the preliminary version presented at ICHEP'04
they were neglected.

\section{THE \babar\ DETECTOR AND DATASET}
\label{sec:babar}

The results presented in this paper are based on data collected
in 1999--2004 with the \babar\ detector~\cite{BABARNIM}
at the PEP-II asymmetric $e^+e^-$ collider~\cite{pep}
located at the Stanford Linear Accelerator Center.  An integrated
luminosity of 198~fb$^{-1}$, corresponding to
218 million \BB\ pairs, was recorded at the $\Upsilon (4S)$
resonance
(``on-resonance'', center-of-mass energy $\sqrt{s}=10.58\ \gev$).
An additional 15~fb$^{-1}$ were taken about 40~MeV below
this energy (``off-resonance'') for the study of continuum background in
which a light or charm quark pair is produced instead of an \UfourS.

The asymmetric beam configuration in the laboratory frame
provides a boost of $\beta\gamma = 0.56$ to the $\Upsilon(4S)$.
Charged particles are detected and their momenta measured by the
combination of a silicon vertex tracker, consisting of five layers
of double-sided silicon microstrip detectors, and a 40-layer central drift chamber,
both operating in the 1.5-T magnetic field of a solenoid.
The tracking system covers 92\% of the solid angle in the center-of-mass frame.

Charged-particle identification is provided by the average
energy loss (\dedx) in the tracking devices  and
by an internally reflecting ring-imaging
Cherenkov detector (DIRC) covering the central region.
A $K/\pi$ separation of better than four standard deviations ($\sigma$)
is achieved for momenta below 3 \gevc , decreasing to 2.5 $\sigma$ at the highest
momenta in the $B$ decay final states. Photons and electrons are detected by a CsI(Tl) 
electromagnetic calorimeter while muons are identified in the magnetic flux 
return system.

\section{ANALYSIS METHOD}
\label{sec:Analysis}

Monte Carlo (MC) simulations~\cite{geant4} of the signal decay mode, of continuum and 
\BB\ backgrounds are used to establish the event selection
criteria. We make several particle identification requirements to ensure the
identity of all signal pions. For the bachelor charged track we require an 
associated DIRC Cherenkov angle between $-2\,\sigma$ and $+5\,\sigma$ from the 
expected value for a pion. A $B$ meson candidate is characterized kinematically by the energy-substituted 
mass $\mes = \sqrt{(\half s + \pvec_0\cdot \pvec_B)^2/E_0^2 - \pvec_B^2}$ and
energy difference $\DE = E_B^*-\half\sqrt{s}$, where the subscripts $0$ and
$B$ refer to the initial \UfourS\ and to the $B$ candidate  in the lab-frame, respectively,
and the asterisk denotes the \UfourS\ frame.
We require $|\DE|\le0.2$ GeV and $5.25\le\mes\le5.29\ \gevcc$.  We select \aunop\  candidates with the
following requirement on the invariant mass: $0.8<m_{a_1}<1.8$ \gevcc .
 The intermediate dipion state is required to have an invariant mass between  0.51 and 1.1 \gevcc .
The momentum of \aunop\ in the
center-of-mass frame is required to be between 2.3 and 2.7 \gevc. To reduce fake $B$ meson
candidates we require p($\chi^2$) $>$ 0.01 for the $B$ vertex fit. The angular
variable $\mathcal{H}_{a_1}$ (cosine of the angle between the direction of the
bachelor $\pi$ and the flight direction of the $B$  in the \auno\ meson rest frame)
is required to be between $-0.85$ and $0.85$ to suppress combinatorics.

To reject continuum background, we make use
of the angle $\theta_T$ between the thrust axis of the $B$ candidate and
that of
the rest of the tracks and neutral clusters in the event, calculated in
the center-of-mass frame.  The distribution of $\cos{\theta_T}$ is
sharply peaked near $\pm1$ for combinations drawn from jet-like $q\bar q$
pairs and is nearly uniform for the isotropic $B$ meson decays; we require
$|\cos{\theta_T}|<0.65$. The remaining continuum background is modelled from 
``off-resonance'' data. 
We use Monte Carlo simulations of \BzBzb\ and \BpBm\ decays
to look for \BB\ backgrounds, which can come from both charmless and charm decays. 
We find that the decay mode $B^0 \rightarrow D^- \pi^+$, with 
$D^- \rightarrow K^+ \pi^- \pi^-$ and $D^- \rightarrow K^0_S \pi^-$, is the only
significant background. It is included in the maximum likelihood fit. Final results 
have been corrected for a small background contribution due to charmless decays. 

We use an unbinned multivariate maximum-likelihood fit to extract
the signal yields for \\
$B^0\rightarrow  a_1^+(1260) \pi^-$.
The likelihood function incorporates five variables.
We describe the $B$ decay kinematics using: 
\DE, \mes, $m_{a_1}$, a Fisher discriminant \xf\ , and an angular variable A.
The Fisher discriminant combines four
variables: the angles in the \UfourS\ frame of the $B$ momentum and $B$ thrust axis with respect to the beam axis, and
the zeroth and second angular moments $L_{0,2}$ of the energy flow
around the $B$ thrust axis.  The moments are defined by
\begin{equation}
  L_j = \sum_i p_i\left|\cos\theta_i\right|^j,
\end{equation}
where $p_i$ is the momentum of particle $i$, $\theta_i$ is the angle 
between the direction of particle $i$ and the trust axis of the B candidate and the sum
excludes tracks and clusters used to build the $B$ candidate.
We have used an angular variable A in order to distinguish
\appim~ from~ \adueppim ~and ~\pipppim. If X is our
resonance $a_1(J^P=1^+)$,  $a_2(J^P=2^+)$ or $\pi(1300)(J^P=0^-)$ that
decays into three pions, we evaluate in the X meson rest frame the cosine
of the angle between the normal to the plane of the three pions and the
flight direction of the bachelor pion. Since we have on average 1.5 B
candidates per event, we choose the best one using a $\chi^2$ quantity
computed with the $\rho$ mass.
 Since the maximum correlation between the observables in the selected data
is 4\%, we take the probability density function (PDF) for each event to be a 
product of the PDFs for the separate observables.
The product PDF for event $i$ and hypothesis $j$, where
$j$ can be signal (3 types), continuum background or \BB\ background, is given by

\begin{equation}
{\cal P}^i_{j} =  {\cal P}_j (\mes) \cdot {\cal  P}_j (\DE) \cdot
 { \cal P}_j(\xf) \cdot {\cal P}_j (m_{a_1}) \cdot {\cal P}_j (A) .
\end{equation}

There is the possibility that a track from a signal
candidate is exchanged with a track from the rest of the event. We call these events
``self-cross-feed'' (SCF) events. The fraction
of SCF events with respect to the total number of signal events for each type $k$ of signal, $f_{SCF_k}$,
is fixed to the value found with  Monte Carlo signal events (26\%). The likelihood function for the event $i$ is defined as :
\begin{equation}
{\cal L}^{i} = \sum_{k=1}^3 \left(n_k(1-f_{SCF_k}) {\cal P}_{k}^i + n_k f_{SCF_k} {\cal P}_{SCF_k}^i\right)
               +n_{q\bar{q}}{\cal P}_{q\bar{q}}^{i} +
               n_{B\bar{B}1}{\cal P}_{B\bar{B}1}^{i}+ n_{B\bar{B}2}{\cal P}_{B
\bar{B}2}^{i}\,,
 \end{equation}

\noindent where $n_k (k=1,3) $  is the yield for  \appim, \adueppim,
and \pipppim\ respectively,  $n_{q\bar{q}}$ the
number of continuum
 background events, $n_{B\bar{B}1}$ the number of \BB\ background events $D^-\pi^+$ with  $D^- \rightarrow K^+ \pi^- \pi^-$
and  $n_{B\bar{B}2}$ the number of \BB\ background events $D^- \pi^+$ with $D^- \rightarrow K^0_S \pi^-$.
The extended likelihood function for all events is :
\begin{equation}
{\cal L} = \frac{\exp{(-\sum_j n_{j})}}{N!}\prod_i^{N}\sum_j n_{j} {\cal P}^i_{j
}\,,
\end{equation}

\noindent where $n_{j}$ is the yield of events of hypothesis $j$ found by the
fitter, and $N$ is the number of events in the sample.  The first factor takes
into account the Poisson fluctuations in the total number of events.

We determine the PDFs for signal and \BB\ backgrounds from
MC distributions in each observable.  For the continuum background we establish
the functional forms and initial parameter values of the PDFs with  off-resonance data.  We allow the signal $a_1(1260)$ PDF parameters and the most important $q\bar{q}$ background PDF parameters to float in the final fit.
The distributions of invariant mass of ${a_1(1260)}$, $a_2(1320)$ and 
$\pi(1300)$ in signal events are parameterized as relativistic Breit-Wigner line-shapes with a mass dependent width which takes into account the effect of the mass dependent $\rho$ width.
The \mes\ and \DE\ distributions for signal are parameterized as
double gaussian functions. Slowly varying distributions are
parameterized by linear functions.
The combinatoric background in \mes\ is described by a phase-space-motivated
empirical function~\cite{argus}. We model the
\xf\ distribution using a Gaussian function with different widths above
and below the mean. The A distributions are modelled using Gaussians in \appim and polynomials in \adueppim\ and \pipppim. 

\section{RESULTS}
\label{sec:Physics}
We present the measurement of the branching fraction of the $B$ decay to
\appim, considering \adueppim\ and \pipppim\ as sources of background.
By generating and fitting simulated samples of signal
and background events, we verify that our fitting procedure is working
properly.  We find that the minimum $\ln{\cal L}$ value for the
on-resonance data lies well within the $\ln{\cal L}$ distribution
from these simulated samples. Fits to data show no evidence of \pipppim, 
since a negative yield is obtained for this resonance.
For this reason the \pipppim\ component has been left out in final fits to the yields.

The reconstruction efficiency is obtained from the fraction of signal MC events passing the selection criteria once corrected for a bias detected in the fit yield. This bias (about 6\%) is determined from fits to simulated samples, each equal in size to the data and containing a known number of signal MC events combined with events generated from the background PDFs. 

\begin{figure}[htp]

 \begin{minipage}{\linewidth}
  \begin{center}
\vspace{1.0cm}
   \includegraphics[bb=85 155 535 605 ,angle=270,scale=0.30]{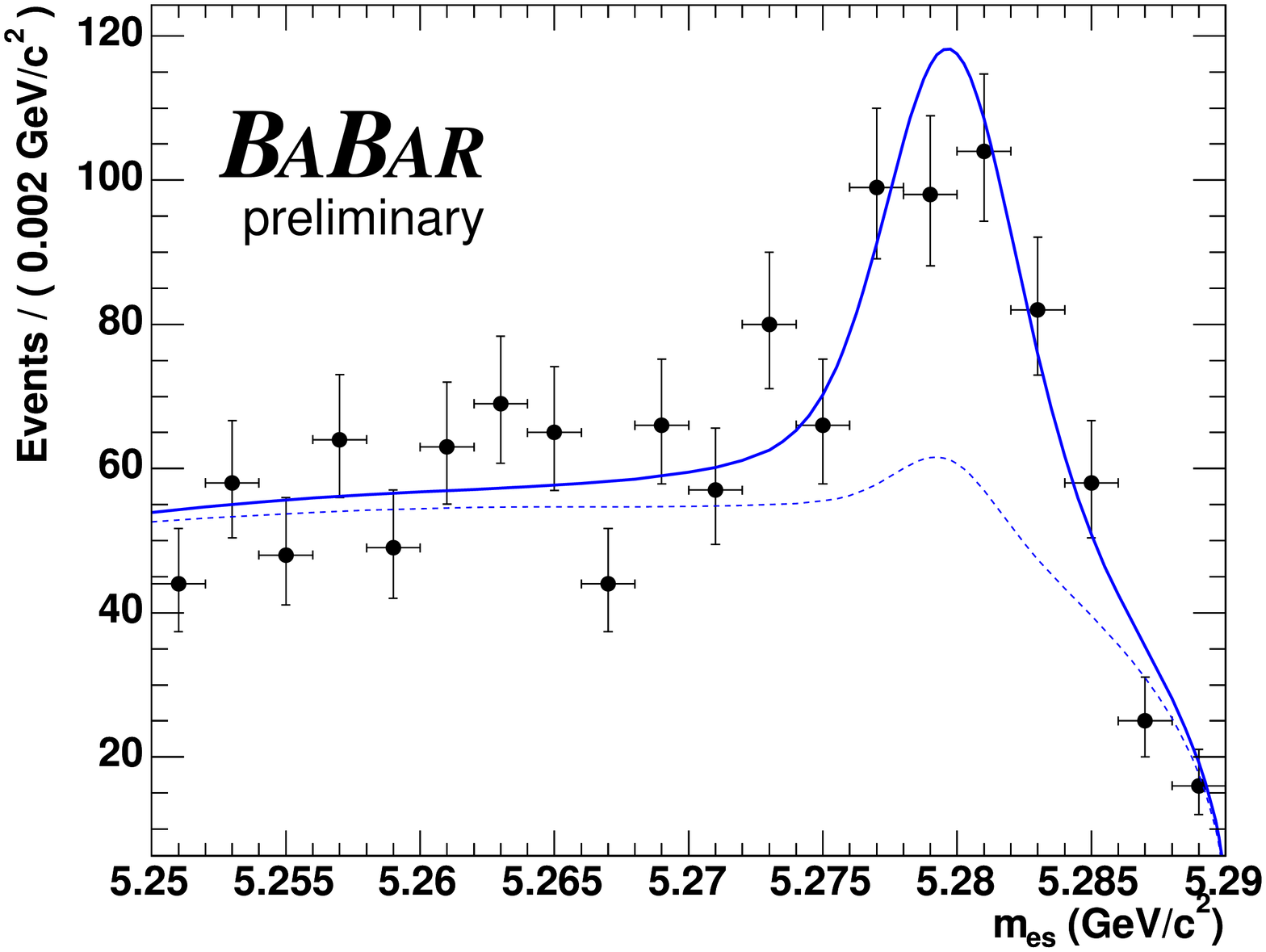}
   \hspace{3cm}
   \includegraphics[bb=85 155 535 605 ,angle=270,scale=0.30]{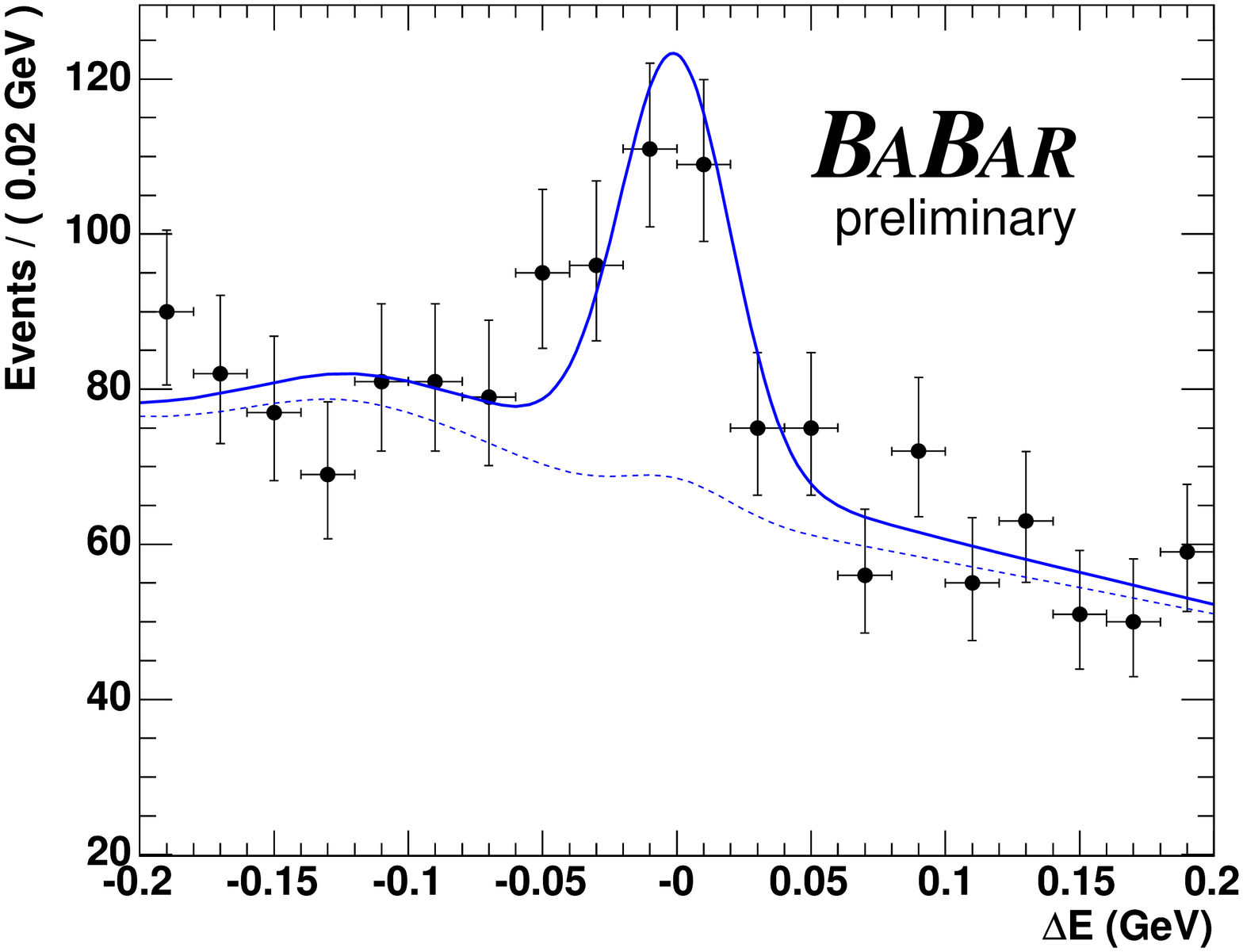}
\vspace{1.5cm}
 \hspace{3cm}
   \includegraphics[bb=85 155 535 605 ,angle=270,scale=0.30]{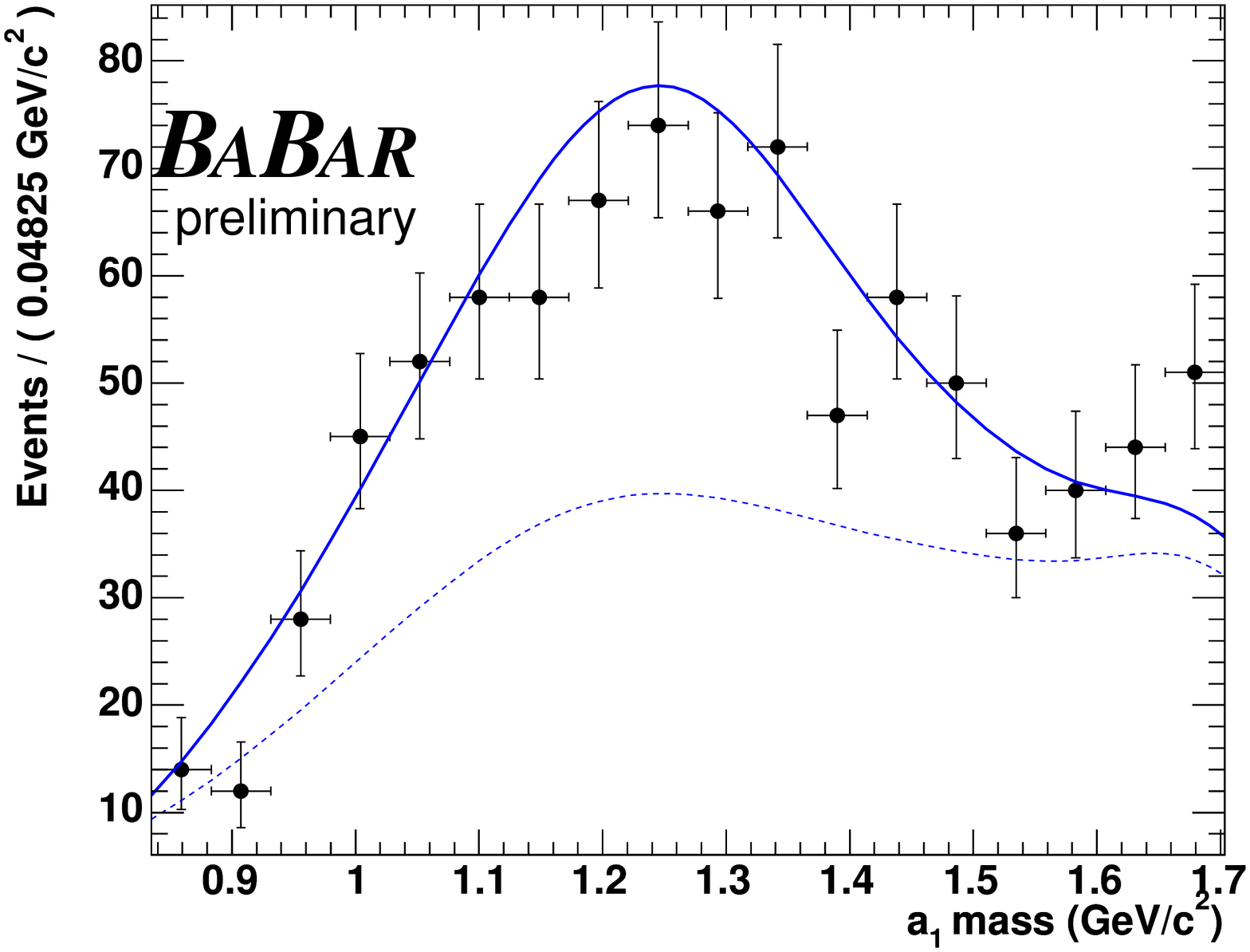}
\vspace{-1.5cm}
   \hspace{3cm}
 \vspace{0.5cm}
   \includegraphics[bb=85 155 535 605 ,angle=270,scale=0.30]{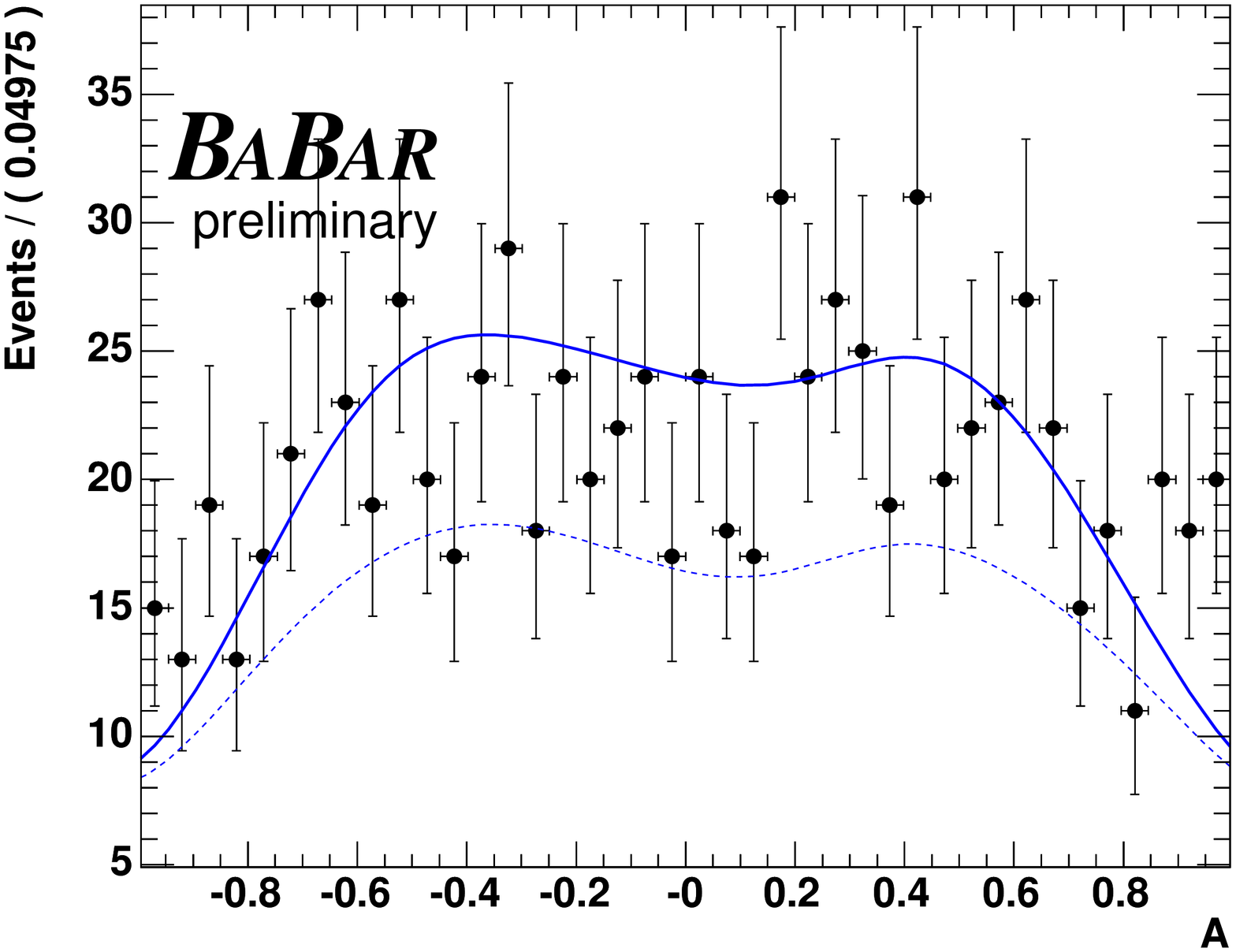}
\end{center}
  \end{minipage}
\vspace{1.cm}
  \caption{ Projections of \mes(a), \DE(b), $a_1$ mass(c) and A(d) for \appim.
Points with errors represent data, dotted lines  the
background from continuum and \BB\ combined, solid curves the full fit functions.
These plots are made with a cut on the signal likelihood and thus do not
show all events in the data sample.}
\label{fig:projections}
\end{figure}

\begin{table}[htb]
\label{tab:pippo}
\begin{center}
\vspace*{-0.2cm}
\hspace*{-1.0cm}
\begin{tabular}{|lc|}
\hline\hline
Quantity    &\appim\  \\
\hline
Signal yield            &$867 \pm 85$ \\
Reconst. $\epsilon$ (\%) &$19.8$\\
$\prod\calB_i$ (\%)      & $50$  \\
\hline
Stat. sign. ($\sigma$)   & $18.5$    \\
${\cal B}(\times 10^{-6})$      &$40.2 \pm 3.9 \pm 3.9$\\
\hline\hline
\end{tabular}
\end{center}
\caption{Final fit results in $B^0\rightarrow\appim$. Fitted signal yield, 
  the final reconstruction efficiency ($\epsilon$), the 
  daughter branching fraction product, the statistical significance, 
  and the central value of the branching fraction with statistical and systematic errors.}
\end{table}
The fitted values of the \auno\ parameters are: $m_{\aunob}=1.22 \pm 0.02$ \gevcc\ and $\Gamma_{\aunob}=0.423 \pm 0.050$ \gevcc\ .
In Table 1 we show the results of the fits for on-resonance data. 
The statistical error on the number of events
is taken to be the change in the central value when the quantity
$-2\ln{\cal L}$ changes by one unit. The statistical significance is
taken as the square root of the difference between the value of
$-2\ln{\cal L}$ for zero signal and the value at its minimum.
In Fig.\ \ref{fig:projections}\ we show the \mes, \DE, $m_{a_1}$ and A 
projections made by selecting events with a signal likelihood (computed 
without the variable shown in the figure) exceeding a threshold that 
optimizes the expected sensitivity.

\section{SYSTEMATIC STUDIES}
\label{sec:Systematics}

Most of the systematic errors on the yields that arise from uncertainties in the
values of the PDF parameters have already been incorporated into the overall
statistical error, since they are floated in the fit.  
We determine the sensitivity to the other parameters of the signal PDF
components by varying these within their uncertainties.  The result is
shown in the first row of Table 2. This is the only
systematic error on the fit yield; the other systematics apply to either
the efficiency or the number of \BB pairs in the data sample.

The uncertainty in our knowledge of the efficiency is found
to be 0.8$N_t$\%, where $N_t$ is the number of signal tracks.
We estimate the uncertainty in the number of
\BB pairs to be 1.1\%. The fitting algorithm introduces 
a systematic bias of 2.8\%, which was found from fits to
simulated samples with varying background populations.  Published world
averages~\cite{PDG2004}\ provide the $B$ daughter branching fraction
uncertainties. The systematic error from $a_1(1260) K$ cross-feed background
is estimated to be 1.4\%, while the systematic error due to SCF is found to be 3.5\%.
 We also take into account systematic differences between data and MC for the 
\costhr\ selection (1.8\%) and the possibility of interference between the $a_1$ and $a_2$ amplitudes (4\%). 
The values for each of these contributions are given in Table 2.

\clearpage
\begin{table}[htbp]
\label{tab:syst}
\begin{center}
\begin{tabular}{|l|c|}
\hline\hline
Quantity & $a_1^+ \pi^- $\\
\hline
Fit yield              & $6.2$   \\
Fit eff/bias           & $2.8$  \\
Track multiplicity     & $1.0$ \\
Tracking eff           & $3.2$ \\
Number \BB\            & $1.1$\\
SCF                    & $3.5$ \\
$a_1 K$ cross-feed     & $1.4$ \\
MC statistics          & $0.6$ \\
\costhr                & $1.8$ \\
$a_1$-$a_2$ Interf.    & $4.0$\\
\hline
Total                  & $9.6$  \\
\hline\hline
\end{tabular}
\end{center}
\caption{Estimates of the systematic errors (in percent).}
\end{table}

\section{SUMMARY}
\label{sec:Summary}
We have obtained a preliminary measurement of the branching fraction for $B^0$ meson decays to \appim\ with \aunop\ $\rightarrow \pi^+ \pi^+ \pi^-$.
The measured  branching fraction is:

\begin{equation}
\Brapi = \Rapi
\end{equation}
The fitted values of the \auno\ parameters are: $m_{\aunob}= 1.22 \pm 0.02$ \gevcc\ and $\Gamma_{\aunob}= 0.423 \pm 0.050$ \gevcc. 

\section{ACKNOWLEDGMENTS}
\label{sec:Acknowledgments}
We are grateful for the 
extraordinary contributions of our \pep2\ colleagues in
achieving the excellent luminosity and machine conditions
that have made this work possible.
The success of this project also relies critically on the 
expertise and dedication of the computing organizations that 
support \babar.
The collaborating institutions wish to thank 
SLAC for its support and the kind hospitality extended to them. 
This work is supported by the
US Department of Energy
and National Science Foundation, the
Natural Sciences and Engineering Research Council (Canada),
Institute of High Energy Physics (China), the
Commissariat \`a l'Energie Atomique and
Institut National de Physique Nucl\'eaire et de Physique des Particules
(France), the
Bundesministerium f\"ur Bildung und Forschung and
Deutsche Forschungsgemeinschaft
(Germany), the
Istituto Nazionale di Fisica Nucleare (Italy),
the Foundation for Fundamental Research on Matter (The Netherlands),
the Research Council of Norway, the
Ministry of Science and Technology of the Russian Federation, and the
Particle Physics and Astronomy Research Council (United Kingdom). 
Individuals have received support from 
CONACyT (Mexico),
the A. P. Sloan Foundation, 
the Research Corporation,
and the Alexander von Humboldt Foundation.


\begin{thebibliography}{99}
\bibitem{Coniugati}
Charge-conjugate reactions are implied throughout this paper.
\end{thebibliography}
\end{document}